\documentclass[preprint,aps,showpacs,floatfix]{revtex4}
\usepackage{graphicx}
\usepackage{amsmath,amssymb,graphicx,epsfig}

\begin{document}

\title{A method to create disordered vortex arrays in atomic Bose-Einstein condensates}

\author{Enik\H{o} J. M. Madarassy}

\affiliation{School of Mathematics, Newcastle University,
Newcastle--upon--Tyne, NE1 7RU, UK}

\date{\today}

\begin{abstract}
We suggest a method to create quantum turbulence (QT) in a trapped atomic Bose-Einstein condensate (BEC). By replacing in the upper half part of our box the wave function, $\Psi$, with its complex conjugate, $\Psi^{*}$, new negative vortices are introduced into the system. The simulations are performed by solving the two-dimensional Gross-Pitaevskii equation (2D GPE). We study the successive dynamics of the wave function by monitoring the evolution of density and phase profile.

\end{abstract}


\maketitle


\section{Introduction}
The aim of this paper is to explore the physics of a turbulent vortex system. Turbulence is studied both in classical and in quantum fields.
\bigskip

A dynamical and statistical description of the classical turbulent flow has been provided earlier by Kolmogorov \cite{Kolmogorov}. The turbulent flow is characterized by random and chaotic three-dimensional vorticity and irregularity. Rapid mixing of momentum, heat and mass is manifested as diffusivity, which is a key characteristic of the turbulent flows. Under the influence of viscosity in the turbulent motion, the kinetic energy is dissipated into heat. Turbulence happens with high Reynolds numbers and is generally anisotropic. 
Simulations using the Gross-Pitaevskii equation (GPE), modelling qualities of both classical and quantum turbulence, have been well studied \cite{Nore,Kobayashi,Parker}.
\bigskip

Key properties of superfluid vortex lines were discovered by Onsager, \cite{Onsager} and later developed by Feynman, \cite{Feynman}. In superfluids like $^4$He or $^3$He, QT was studied with quantized vortices characterized by topological defects \cite{Vinen}. Configurations of quantized vortices were investigated in superfluid $^4$He \cite{Donnelly}, which can be grouped into two types \cite{Saffman}: ordered vortex arrays \cite{Yarmchu} and disordered vortex tangles \cite{Vinen1,Schwarz}. 
Disordered vortex tangles appear, when helium is made turbulent under thermal counterflow velocity or using grids or propellers \cite{Barenghi}, 
for example by moving a grid at $T=0.01T_{\lambda}$ where, $T_{\lambda}=2.17$ \cite{Davis} or by a vibrating-wire resonator in B-phase superfluid $^3$He ($^3$He-B) \cite{Fisher}. 
Large scale turbulence of quantized vortices was studied in superfluid $^3$He-B \cite{Vinen} and $^4$He \cite{Donnelly}.
The disadvantage of turbulence in BEC is the small system size and the relatively low number of vortices. On the other hand, the advantage is the relatively good visualization in detail of the individual vortices.
\bigskip

In magnetically or optically trapped rotating atomic Bose-Einstein condensates (BECs), vortex lattices with quantized vortices along the rotation axis are formed. The density and phase of BECs can be directly observed. The rotational motion is sustained by quantized vortices and is quantized in units of $\kappa = q (h/m$), where $q$ is integer, $m$ is the particle mass and $h$ is the Planck constant. For $q > 1$ the system is unstable . 
\bigskip

The circulation around each superfluid vortex filament is fixed by the condition: \\
$\int_{C} \mathbf{{v}_{s}} d{\bf r}= \kappa$ where, $C$ is a circular path around the axis of the vortex and $\mathbf{v}_{s}$ is the local velocity flow.
\bigskip

Quantized vortices are characterized by singularities in the phase and by associated holes in the density. Vortex lattices were first realized experimentally by the groups of Madison \cite{Madison} and Abo-Shaeer \cite{Abo-Shaeer}.
The corresponding simulations were made for example, by the group of Tsubota \cite{Kasamatsu}.
\bigskip

Similarities between QT and classical turbulence (CT) were observed experimentally \cite{Stalp,Finne}. A realizable study of QT was presented by using numerical simulations of the GPE in trapped BEC, and by combining rotations around two axes \cite{Kobayashi1}. The kinetic energy, $E_\mathrm{kin}$, can be divided into a compressible part, $E_\mathrm{kin}^{c}$, due to the sound waves, and into an incompressible part, $E_\mathrm{kin}^{i}$, due to the vortices \cite{Kobayashi}. 
It was shown that, the spectrum of the incompressible kinetic energy obeys the Kolmogorov law and the energy flux becomes constant value. On the other hand, the method we present here is a pure mathematical model, which can give ideas and inspirations to the experimental investigations. With this method a superfluid analogy of classical 2D turbulence was created in a 2D system. 
\bigskip

The wave function or the order parameter of the solution of GPE is a complex number, $\psi$, corresponding to the amplitude of the particle to have a given position, $\mathbf{r}$ at any given time, $t$. 
The condensate wave function, $\psi$ in a complex plane or Argand diagram is observed as a positive vector with the real part, $\mathrm{Re}~\psi$ and the imaginary part, $\mathrm{Im}~\psi$. The real and imaginary parts are regarded as independent quantities. 
\bigskip 

The condensate density is defined as $n(\mathbf{r},t)=\vert \psi(\mathbf{r},t) \vert^{2}$, while the phase of the condensate as $S=\tan^{-1}\left[{{\rm Im}~\psi(\mathbf{r},t)}/{{\rm Re}~\psi(\mathbf{r},t)}\right]$. Using the Madelung transformation, that is to say the polar form of $\psi$ in terms of the superfluid density and macroscopic phase, we obtain $\psi(\mathbf{r},t)=\sqrt{\vert \psi(\mathbf{r},t)\vert^2}e^{iS(\mathbf{r},t)}=\sqrt{n(\mathbf{r},t)}e^{iS(\mathbf{r},t)}$. 
The superfluid local velocity flow is given by $\mathbf{v}_{s}(\mathbf{r},t)=(\hbar / m)\nabla S(\mathbf{r},t)$. 
\bigskip 

The complex conjugate of a complex number is given by changing the sign of the imaginary part. The complex conjugate of $\psi(\mathbf{r},t)$ is $\psi^{*}={\rm Re}~\psi-i~{\rm Im}~\psi$ and similarly $\psi^{*}(\mathbf{r},t)=\sqrt{\vert \psi(\mathbf{r},t)\vert^{2}}e^{-iS(\mathbf{r},t)}$ = $\sqrt{n(\mathbf{r},t)}e^{-iS(\mathbf{r},t)}$.
In the complex plane, $\psi^{*}$ is symmetric about the real axis and $\psi + \psi^{*}$ and $\psi \cdot \psi^{*}$ are real numbers. If a complex number supplies a solution to a problem, so its conjugate does too.
\bigskip 

\section{Theory: The Gross-Pitaevskii equation} 
To explore the dynamics of superfluid vortices at nonzero temperatures and the density and phase profile of a rotating condensate, we solve numerically the following 2D GPE for the condensate macroscopic wave function, $\psi(\mathbf{r},t)$:

\begin{equation}
(i-\gamma)\hbar \frac{\partial \Psi}{\partial t} = \left[-\frac{\hbar^{2}}{2m}\nabla_{\perp}^{2}
+V_\mathrm{trap}+g_{2D}N \vert \Psi\vert^{2}-\mu-\Omega L_{z}\right]\Psi,
\label{eqn:GP}
\end{equation}
where, $m$ is the atomic mass, $\emph{a}$ is the $s$-wave scattering length, $\Omega$ is the angular frequency of rotation about the $z$-axis and $L_{z}=i\hbar(x\partial_{y}-y\partial_{x})$ is the angular momentum operator. 
The rotation frequency, $\Omega$ is large enough, so several vortices are present in the field, $\psi$ forming a regular array. The phenomenological damping parameter is $\gamma=0.01$, 
which models the interaction of the condensate with the surrounding thermal cloud \cite{Tsubota,Madarassy}. In 2D due to the enough strong confinement along the $z$-axis the coupling constant becomes \cite{Regnault}:

\begin{equation}
g_\mathrm{2D} = \sqrt{32 \pi} \hbar \Omega \frac{a l^{2}}{l_{z}}.
\label{eqn:g2d}
\end{equation}
By using the following formulas, $l^{2}=\hbar/2m \Omega$ and $l_{z}^{2}=\hbar/m \omega_{z}$ and by multiplying the numerator and the denominator by $\omega_{z}$, we obtain finally:

\begin{equation}
g_\mathrm{2D} = 2 \sqrt{2 \pi} \hbar a l_{z} \omega_{z},  
\label{eqn:g2d_2}
\end{equation}
where, $\Omega$ is the angular velocity along the $z-axis$, $l$ is the magnetic length, $l_{z}$ is the characteristic length of the z-axis oscillator and $\omega_{z}$ is the characteristic frequency of the trapping $z$ potential. In dimensional form the chemical potential, $\mu$ is defined as:

\begin{equation}
\mu= \omega  \sqrt{\frac{Ngm}{\pi}},
\label{eqn:MuI}
\end{equation}
In imaginary time, excitations are damped and both $\psi(\mathbf{r},t)$ and $\mu$ converge to a stationary solution supplying exact initial conditions for time-dependent solutions. In 2D the harmonic trapping potential is defined by:

\begin{equation}
       V_\mathrm{trap}(x,y)=\frac{1}{2}m\omega_{\perp}^{2}\left(x^{2}+y^{2}\right),
\label{eqn:VTR}
 \end{equation}
here, the radial trap frequency is $\omega_{\perp}$. 
\bigskip
The ground state properties of disk-shaped BECs ($\omega_{z} \gg \omega_{\perp}$) are characterized by only one parameter \cite{Munoz}: 

\begin{equation}
K_{2} \equiv \frac{N a}{\lambda l_{z}},
\label{eqn:PK2}
\end{equation}
here, $N$ is the number of particles, $a$ is the $s$-wave scattering length, $l_{z}$ is the axial oscillator length, and $\lambda \equiv \omega_{z}/ \omega_{\perp} \gg 1$, is the trap aspect ratio. With the help of $\lambda$, $l_{z}$ and $l_{\perp} = \sqrt{\hbar/m \omega_{\perp}}$ (which is the radial oscillator length) expressions, we obtain the final form of $K_{2}$:

\begin{equation}
K_{2}= N a l_{z}^{3} / l_{\perp}^{4}.
\label{eqn:PK2a}
\end{equation}
The perturbative regime (quasi-2D regime) corresponds to $K_{2} \ll 1$, while for the Thomas-Fermi (TF) regime we have the condition, $K_{2} \gg 1$. In the perturbative regime, the axial wave function coincides essentially with the ground state (Gaussian) wave function of the corresponding (z) harmonic oscillator. 
On the other hand, in the TF regime the axial wave function is essentially a TF wave function.
\bigskip

Thus, only when $K_{2} \ll 1$ one can assume that, the wave function along the $z$-axis is the ground state of the harmonic potential. This only occurs when, $N$ and hence the nonlinear mean-field interaction term in the equation of motion is small enough or $\lambda$ is large enough.
There exists a direct relation between $K_{2}$ and the parameter, $a l_{z} n_{2}(\mathbf{0})$ where, $n_{2}(\mathbf{r}_{\perp},t)$ is the local condensate density per unit area characterizing the radial configuration \cite{Munoz1}. 

\begin{equation}
 n_{2}(\mathbf{r}_{\perp},t) \equiv  N \int dz \vert \Psi(\mathbf{r}_{\perp},z,t)\vert^{2}.         
\label{eqn:n2}
\end{equation}

In fact, for $K_{2} \ll 1 \Rightarrow a l_{z} n_{2} \ll 1$ and for $K_{2} \gg 1 \Rightarrow a l_{z} n_{2} \gg 1$. 
The (axial) local chemical potential, $\bar{\mu}_{z}=\mu_{z}/\hbar \omega_{z} $ \cite{Munoz1} is used in \cite{Munoz2} to obtain an effective 2D equation of motion for disk-shaped condensates. In the quasi-2D mean-field regime, $K_{2} \ll 1 \Rightarrow a l_{z} n_{2} \ll 1$. This effective equation reduces to Eq.~\ref{eqn:GP}, where $N \vert \Psi(\mathbf{0}) \vert^{2}=n_{2}(\mathbf{0}) \sim n_{2}$. 
One cannot know the value of $n_{2}$ in advance, therefore in practice $K_{2}$ is the relevant parameter, which is a global parameter. In the TF regime, $K_{2} \gg 1 \Rightarrow a l_{z} n_{2} \gg 1$. The effective 2D equation \cite{Munoz2} reduces to:

\begin{equation}
(i-\gamma) \hbar \frac{\partial \Psi}{\partial t} = \left[-\frac{\hbar^{2}}{2m} \nabla_{\perp}^{2}
+V_\mathrm{trap}+ \hbar \omega_{z} \left( \frac{3 \pi}{\sqrt{2}} a l_{z} N \vert \Psi \vert^{2} \right)^{2/3}\right]\Psi.
\label{eqn:GP1}
\end{equation}
In the above equation, we are assuming $\Psi$ to be normalized to unity. Both Eq.~\ref{eqn:GP} and Eq.~\ref{eqn:GP1} are the correct limits of the underlying three-dimensional GPE.
\bigskip   

It is also necessary that, the typical time scale of the radial motion, ($\Delta_{t}$) to be much larger (i.e. slower) than the time scale of axial motion (which is of the order of $\omega_{z}^{-1}$).
This is necessary, in order for the adiabatic approximation to be valid, and as a result radial and axial motions to be separable \cite{Munoz2}.
\bigskip   

For stationary problems, $ \Delta_{t} \sim \infty$, and for collective oscillations, $\Delta_{t} \sim \omega_{\perp}^{-1}$, and thus the condition, $\lambda \equiv \omega_{z}/\omega_{\perp} \gg 1$ guarantees the fulfillment of the adiabatic approximation.
Connect with Eq.~\ref{eqn:GP}, there is no problem, since $\Delta_{t} \sim (\omega_{z} a l_{z} n_{2})^{-1} \gg \omega_{z}^{-1}$, which is a consequence of the fact that, in this case $a l_{z} n_{2} \ll 1$.
\bigskip   

The total energy, $E_\mathrm{tot}$, can be identified with the sum of three energies:
\begin{equation}
      E_\mathrm{tot}=E_\mathrm{kin}+E_\mathrm{int}+E_\mathrm{trap},
\label{eqn:TotE}
\end{equation}
where, the kinetic energy, $E_\mathrm{kin}$, the internal energy, $E_\mathrm{int}$, and the trap energy, $E_\mathrm{trap}$ are given respectively by:
\begin{eqnarray}
E_\mathrm{kin}(t)&=&\int \frac{\hbar^{2}}{2m} \left(\sqrt{\rho({\bf x}, t)}{\bf v}({\bf x}, t) \right) ^{2} d^{2}\mathbf{r},\label{eqn:KE}\\
E_\mathrm{int}(t)&=& \frac{1}{2} \int g \left(\rho({\bf x}, t)\right)^{2} d^{2}\mathbf{r},\label{eqn:IE}\\
E_\mathrm{trap}(t)&=&\int \rho({\bf x}, t)V_\mathrm{tr}(\mathbf{x})  d^{2}\mathbf{r}.\label{eqn:TrE}
\end{eqnarray}
In the kinetic energy expression, the main contribution to he energy comes from the phase gradients. The contribution from the density gradients is neglected due to its small value.
\bigskip

It is useful to scale the GPE in dimensionless units. We use harmonic-oscillator units (h.o.u.) in the whole paper \cite{Ruprecht}, where the units of time, length and energy are: $\omega^{-1}_{\perp}$, $\sqrt{\hbar/m \omega_{\perp}}$ and $\hbar\omega_{\perp}$ respectively, so that:
\bigskip

\begin{equation}
\left (i-\gamma\right )\frac{\partial\psi}{\partial t}=\left [-\frac{1}{2}\nabla_{\perp}^{2}+V_\mathrm{trap}+C \vert \psi \vert ^{2}-\mu-\Omega L_{z}\right ]\psi,
\label{eqn:2D_GPE}
\end{equation}
where,

\begin{equation}
V_\mathrm{trap}=\frac{1}{2}\left( x^{2}+y^{2}\right),
\end{equation}
and the effective coupling constant, which is an important characteristic parameter of the 2D system becomes:

\begin{equation}
C \equiv \frac{g_{2D} N}{l_{\perp}^{2} \hbar \omega_{\perp}},
\label{eqn:C1}
\end{equation}
where, $g_{2D}$ is expressed by Eq.~\ref{eqn:g2d} and by Eq.~\ref{eqn:g2d_2}. Using the expression from Eq.~\ref{eqn:g2d_2} together with the expression of $l_{\perp}$, $l_{\perp} = \sqrt{\hbar/m \omega_{\perp}}$ and then by substituting $m \omega_{z}/ \hbar$ by $1 /l_{z}^{2}$, 
we obtain first:

\begin{equation}
C = 2 \sqrt{2 \pi} \frac{aN}{l_{z}},
\label{eqn:C2}
\end{equation}
and later by substituting $N a / l_{z} \lambda^{2}$ by $K_{2}$, we obtain in 2D the final form of the effective coupling constant:

\begin{equation}
C = 2 \sqrt{2 \pi} \lambda^{2} K_{2}.
\label{eqn:C3}
\end{equation}
Here, $N$ represents the number of atoms per unit length along $z$. Throughout this paper we use, $C=1400$ in these calculations. In order for Eq.~\ref{eqn:2D_GPE} to be valid, we need $K_{2} \ll 1$. Say for example, $K_{2}=0.1$:

\begin{equation}
K_{2}= \frac{C}{\sqrt{8 \pi} \lambda^{2}} \ll 1, 
\end{equation}
which, implies that:

\begin{equation}
\lambda^{2} \gg  \frac{C}{\sqrt{8 \pi}},      
\end{equation}
and 

\begin{equation}
\omega_{z}^{2} \gg \frac{C}{\sqrt{8 \pi}}\omega_{\perp}^{2}.
\end{equation}
\bigskip

We would like to present an example for the experimental realization. For a $^{87}$Rb condensate, the $s$-wave scattering length is $a=5.82$~nm, and the axial oscillator length is:

\begin{equation}
l_{z} \equiv \sqrt{\frac{\hbar}{m \omega_{z}}}=\frac{10.78}{\sqrt{\omega_{z}/2 \pi}} \mu m. 
\end{equation}
Substituting these values for $a$ and $l_{z}$ in Eq.~\ref{eqn:C2}, we obtain:

\begin{equation}
C=2.7066 \times 10^{-3} \sqrt{\omega_{z}/2 \pi}N,
\end{equation}
and using, $C=1400$ one finds:

\begin{equation}
N = \frac{517254.03}{\sqrt{\omega_{z}/2 \pi}}.
\end{equation}
Considering typical values, as for instance, $\omega_{z}/2 \pi=600$~Hz and $\omega_{\perp}/2 \pi=10$~Hz, one has a condensate with 
$N=21117 \simeq 21100$ particles and:

\begin{equation}
\lambda \equiv \omega_{z}/ \omega_{\perp}=60,
\end{equation}
and

\begin{equation}
K_{2}=\frac{C}{\sqrt{8 \pi} \lambda^{2}}= 0.0776 \simeq 0.08.
\end{equation}
Moreover, one can easily estimate (rather accurately) the radius, $R$ and the chemical potential, $\mu$ of the ground state 
of this condensate \cite{Munoz}:

\begin{equation}
\frac{\mu}{\hbar \omega_{z}}=\frac{1}{2}+ \left( 2 \sqrt{2/ \pi}K_{2} \right)^{1/2}=0.852,
\end{equation}
and 

\begin{equation}
\frac{R}{l_{\perp}}=\sqrt{\lambda} \left( 8 \sqrt{2/ \pi}K_{2} \right)^{1/4}=6.498.
\end{equation}
Using that, $l_{\perp}= \sqrt{\lambda}~l_{z}=\sqrt{60}~10.78/\sqrt{600}=3.41~\mu m$, we find $R=22.15~\mu m$. On the other hand, since $\mu=0.852~\hbar \omega_{z}< \hbar \omega_{z}$,
we conclude that, the condensate is indeed in its axial ground state (quasi-2D regime). However,

\begin{equation}
\frac{\mu}{\hbar \omega_{\perp}} = \lambda \frac{\mu}{\hbar \omega_{z}}=51.12,
\end{equation}
$\Rightarrow \mu \gg \hbar \omega_{\perp}$, which means, that with respect to its radial motion, the condensate is in a Thomas-Fermi 
regime (many radial modes excited). This is also confirmed by the fact that, $R \gg l_{\perp}$.
\bigskip

We consider a strongly interacting 2D BEC in a harmonic trap rotating at the angular frequency, $\Omega$. The numerical calculations are performed with the semi-implicit, Crank-Nicholson method in a square box of size $+/-$ $7/8$ (h.o.u.).
\bigskip

\section{Results}
\begin{figure}[p]
      \epsfig{figure=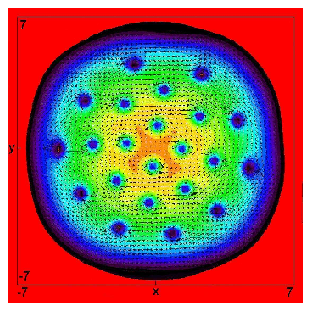,width=0.49\linewidth}
      \epsfig{figure=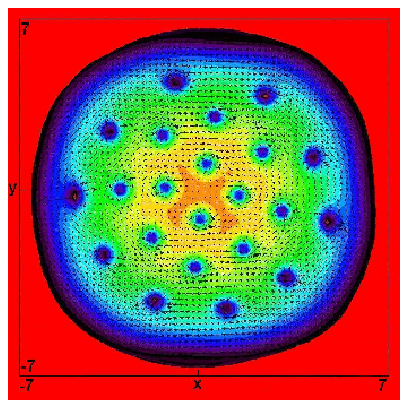,width=0.49\linewidth}
      \caption{For $\Omega= 0.85$, the density profile of the condensate at $t=199$ (left) and $t=200$ (right), after applying the method described in this paper. The vortices rotate in opposite directions in the upper half and in the bottom half part of the condensate. (The squared shape of the condensate depends on the small size and the boundary effect of the box.)}
\label{fig:f1}
 \end{figure}

\begin{figure}[p]
      \epsfig{figure=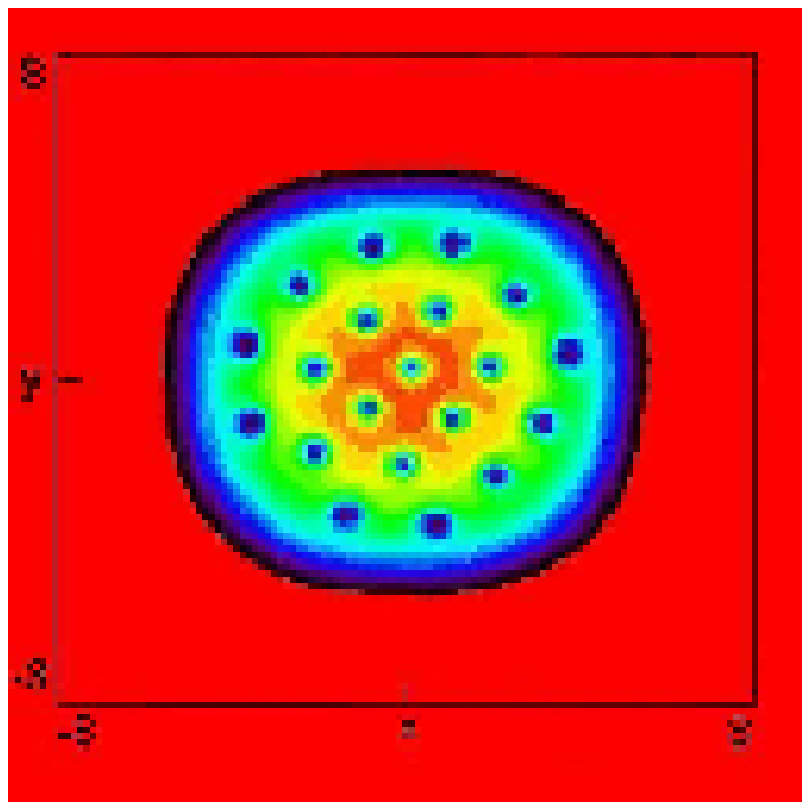,height=4in,angle=0,scale=0.6}
      \epsfig{figure=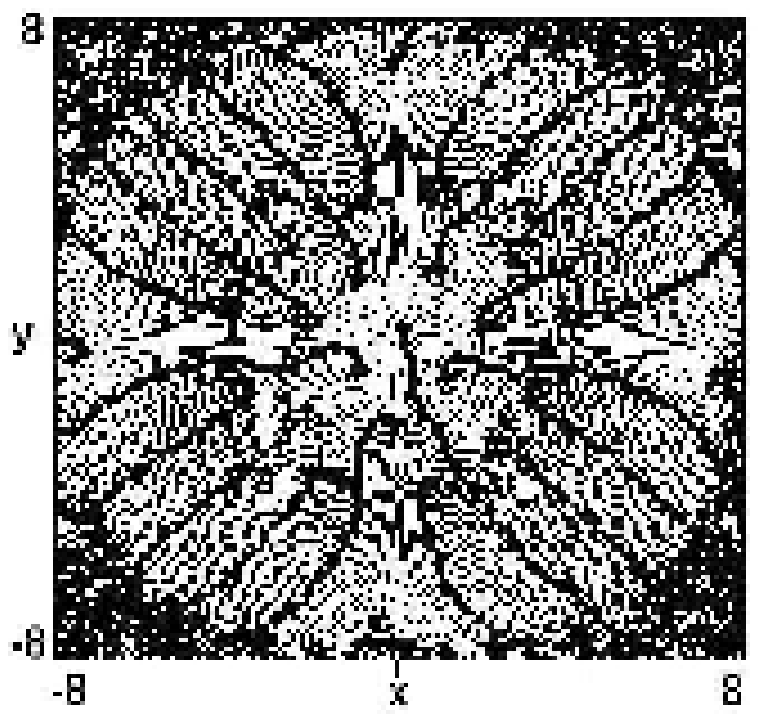,height=4in,angle=0,scale=0.6}
      \caption{The density (left) and phase (right) profiles of the condensate for $\Omega=0.8$ at $t=199$.}
\label{fig:f2}
 \end{figure}

First, we create a non-rotating condensate at $t=0$, and later $\Omega$ is set to $ 0.85/0.8$ to create a stable lattice of $22/20$ vortices in a rotating frame. Physical quantities of the BECs like density and phase can be clearly observed. Arrays of vortices are shown by the density profile of the condensate for $\Omega=0.85$ at $t=199$, see Fig.~\ref{fig:f1} (up) or by the density and phase profile of the condensate for $\Omega=0.8$ at $t=199$, see  Fig.~\ref{fig:f2}. 
Rotation of a superfluid takes place via Abrikosov lattice of quantised vortices, where the rotational velocity profile mimics solid body rotation. 
Away from vortex cores the superfluid is irrotational. 
For vortex lattice with $N$ vortices, the circulation becomes, 
$\Gamma = \int \mathbf{v}_{s} \cdot d \mathbf{l} = N \kappa$ and the vortex density depends only on $\Omega$, as we can see from its expression, $n_{v} = N /A = 2 \Omega / \kappa$. 
Here, $\Omega$ is the angular velocity of the trapping potential and $N$ is the number of vortices. The circulation of the fluid is quantised in units of $\kappa$.
\bigskip

Disordered vortex arrays are produced with the help of the following method: $\Psi$ is instantaneously changed to $\Psi^{*}$ in the upper half part of the box ($y>0$; the center of the box is at $[x,y]=[0,0]$). That means that, the condensate is divided into two equal parts with opposite rotation. The upper part rotates anti clockwise and the bottom part rotates clockwise. These anti-vortices in the upper part of the condensate change their sense of rotation and are visualized by Fig.~\ref{fig:f1} (down). 
\bigskip

\begin{figure}[p]
      \epsfig{figure=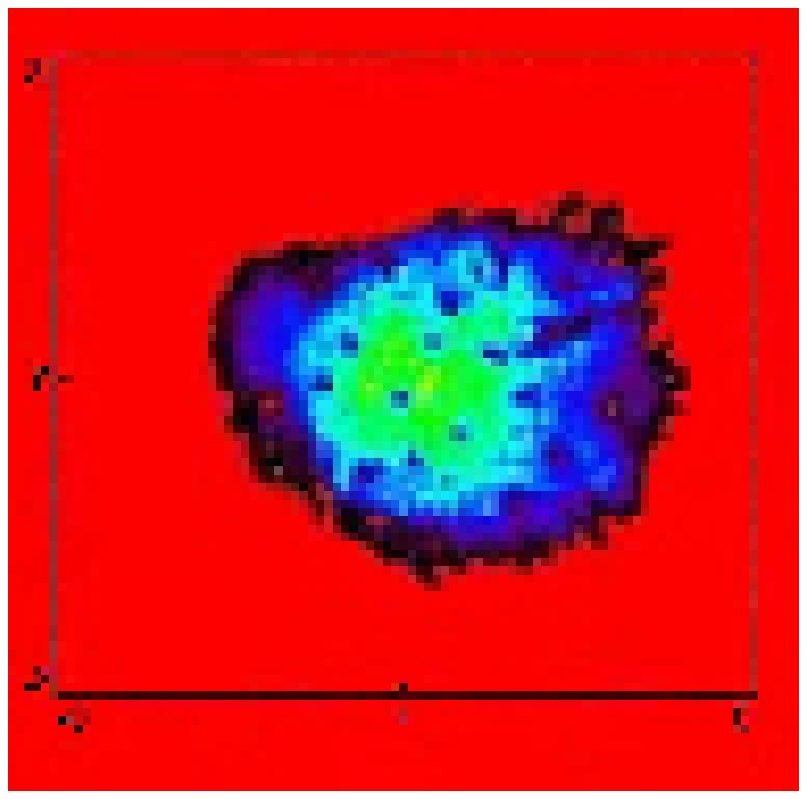,height=4in,angle=0,scale=0.6}
      \epsfig{figure=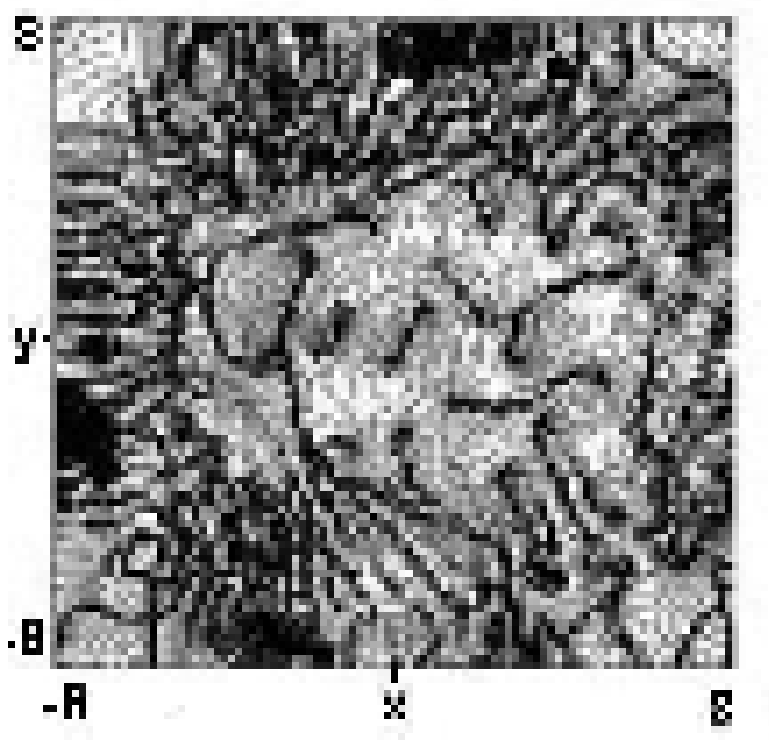,height=4in,angle=0,scale=0.6}
      \caption{The density (left) and phase (right) profiles of the condensate, corresponding to FIG.~\ref{fig:f2} at $t=202.1$.}
\label{fig:f3}
 \end{figure}

\begin{figure}[p]
      \epsfig{figure=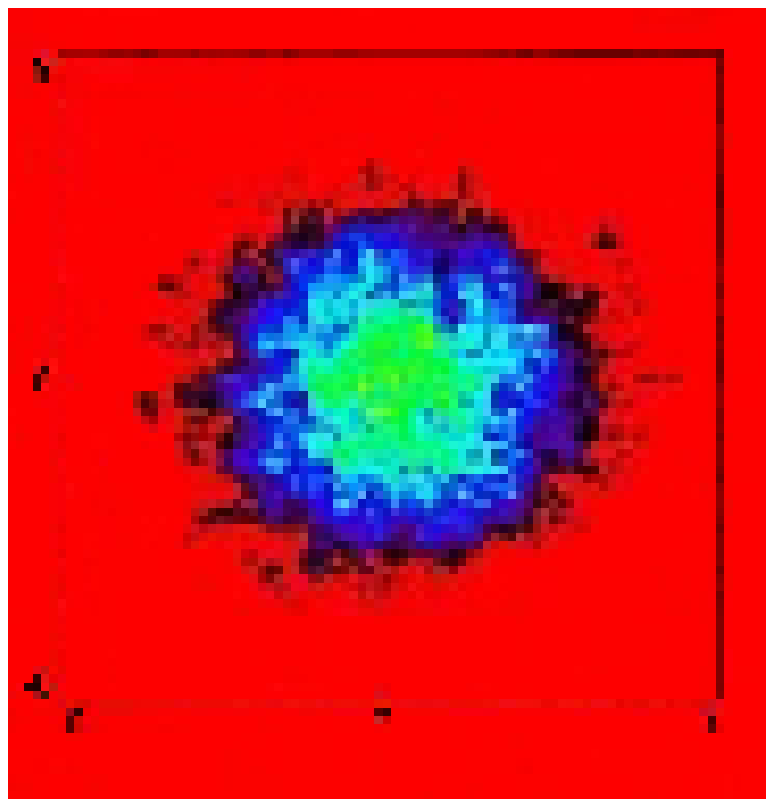,height=4in,angle=0,scale=0.6}
      \epsfig{figure=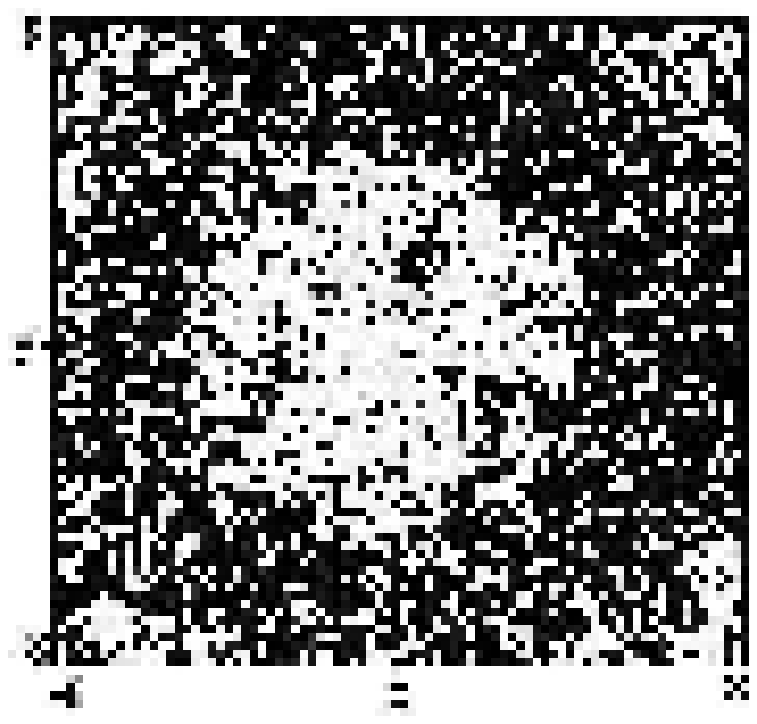,height=4in,angle=0,scale=0.6}
      \caption{The density (left) and phase (right) profiles of the condensate, corresponding to FIG.~\ref{fig:f2} at $t=218.8$.}
\label{fig:f4}
 \end{figure}

This method is a shock for the condensate. First, the largest part of the upper side of the condensate becomes deformed and after some shaking movements, see Fig.~\ref{fig:f3}, the system recovers its circle-like form again with turbulence, shown by Fig.~\ref{fig:f4}. So, the method can thus be used to create a 2D turbulence in the same way as in the classical Onsager vortex gas \cite{Wang}.
\bigskip

\begin{figure}[p]
      \epsfig{figure=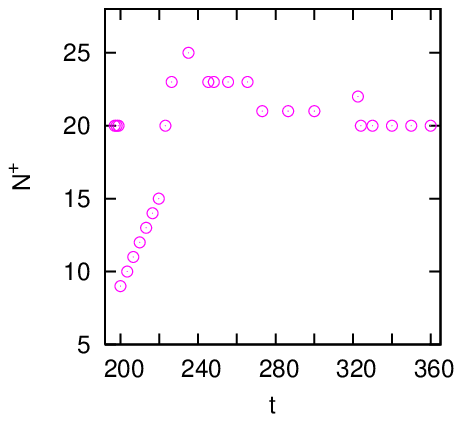,height=3.5in,angle=0,scale=0.6}
      \epsfig{figure=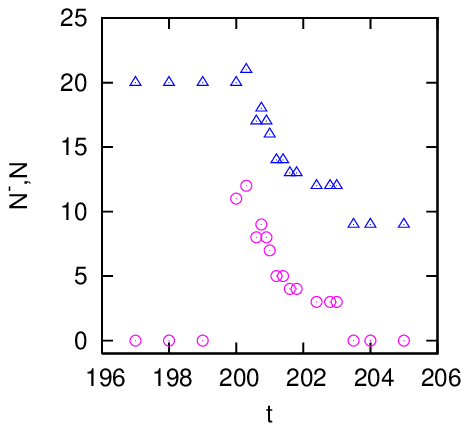,height=3.5in,angle=0,scale=0.6}
      \caption{The number $N^+$ of positive vortices (left), and (right) the total number of vortices ($N$, represented by triangles) along with the number of negative vortices ($N^-$, represented by circles) for the presented method. Here, $\Omega$ was kept at $0.8$ after $t=200$.}
\label{fig:f6}
\end{figure}

After using of this method, we distinguish two different cases: $(\Omega=0.8)$ and $(\Omega=0)$. We analyze the total number, $N$ of vortices, as well as the number $N^{+}$, of positive and $N^{-}$, of negative vortices (see Fig.~\ref{fig:f6} and Fig.~\ref{fig:f7}) under certain conditions, for example keeping $\Omega$ constant or not. 
For $\Omega=0.8$, after a certain transition period, the system enters the state which is the same as the original regular vortex lattice.

\section{Discussion}
First, we create vortex lattice in a rotating trapped BEC. Then, we change the wave function with its complex conjugate by instantaneously reversing the direction of rotation of the atoms in the top half part of the box. As a consequence, in the upper part of the trap, we transform vortices into their anti-vortices rotating in the opposite direction.
\bigskip

\begin{figure}[p]
\centering \epsfig{figure=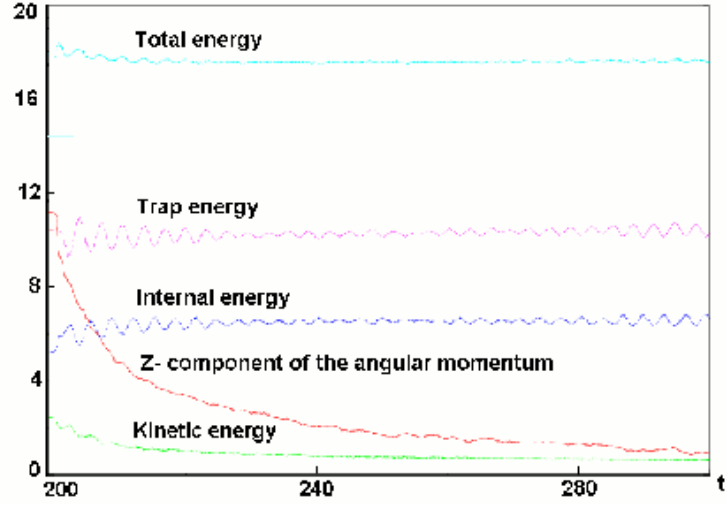,width=0.6\linewidth}
\caption{Different energies and $L_{z}$ for the presented method, as a function of time, $t$ (from bottom to top: kinetic energy (green), $z$-component of the angular momentum (red), internal energy (blue), trap energy (purple) and total energy (cyan)). To keep the anti-vortices inside of the condensate after $t=200$, $\Omega$ was set to $0$.}
\label{fig:f5}
\end{figure}

Thus, we suggest a method to create a system of positive and negative vortices in a disk-shaped condensate, which is divided into two equal parts rotating in opposite directions, as shown in Fig.~\ref{fig:f1} (down). To prepare the right conditions for turbulence, at $t=200$, the rotation frequency, $\Omega$ was set to $0$. 
As a result, the kinetic energy, $E_\mathrm{kin}$ and the $z$-component of the angular momentum, $L_{z}$ tend to zero, (see, Fig.~\ref{fig:f5}).
\bigskip

\begin{figure}[p]
\centering \epsfig{figure=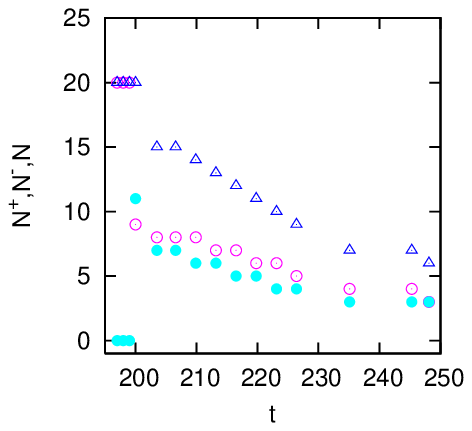,height=4in,angle=0,scale=0.7}
\caption{The total number of vortices ($N$, represented by triangles), the number of positive vortices ($N^+$, represented by circles) and the number of negative vortices ($N^-$, represented, by filled circles), corresponding to FIG.~\ref{fig:f6} for the case of $\Omega=0$ from $t=200$ (when the method was applied).}
\label{fig:f7}
\end{figure}

In that case, the negative vortices remain in the condensate, as seen in Fig.~\ref{fig:f7}. We find that, after applying the recent method, the new anti-vortices interact with the existing positive vortices and contribute to the formation of turbulence, see Fig.~\ref{fig:f4}. 
\bigskip

In two separate cases, we track the number of vortices and anti-vortices by maintaining or discontinuing the trap rotation. If we continue to use $\Omega=0.8$ or 0.85 after the suggested method, the anti-vortices move out to the edge of the condensate, and new vortices come in until the vortices settle into an ordered vortex array. The corresponding total number, $N$, positive, $N^{+}$, and negative, $N^{-}$, numbers of vortices are shown in Fig.~\ref{fig:f6}, when the system will relax to the original vortex array.
\bigskip

On the contrary, in the second case (for $\Omega = 0$) both vortices and anti-vortices coexist and interact with each other to create turbulence. In that case, more negative vortices and more possibilities for annihilations appear, see the suitable plots for $N$, $N^{+}$ and $N^{-}$ in Fig.~\ref{fig:f7}. 
  
\section{Conclusion}
It is instructive to compare this method (call $A$) to create turbulence with complex conjugation of the order parameter in a semi-plan with a different one (call $B$) \cite{Madarassy1}, when the phase was imprinted in upper left quadrant ($x<0$ and $y>0$) and bottom right quadrant ($x>0$ and $y<0$) of the box, or with a method to produce positive and negative vortices by the phase imprinting method (in upper two quadrants), when soliton-like perturbation due to the snake-instability, \cite{Feder} decays into vortex - anti vortex pairs \cite{Madarassy2}. 
In that case, the total number of negative (anti)vortices is not enough to create turbulence. 
\bigskip

In case $B$, the direction of vorticity of quantized vortex-line (in 2D vortex-point) or the sign of circulation does not change. On the other hand, in case $A$ the direction of the condensate does flip. Thus, in $B$, we change the sign of the square root of the density in the Madelung transformatio: $\psi(\mathbf{r},t)=-\sqrt{\vert \psi(\mathbf{r},t) \vert^2}e^{iS(\mathbf{r},t)}=-\sqrt{n(\mathbf{r},t)}e^{iS(\mathbf{r},t)}$. Whereas, in A, we change the sign of the phase: $\psi(\mathbf{r},t)=\sqrt{\vert \psi(\mathbf{r},t)\vert^2}e^{-iS(\mathbf{r},t)}=\sqrt{n(\mathbf{r},t)}e^{-iS(\mathbf{r},t)}$.  
\bigskip

Another important difference between these two methods is that, in the case $B$ soliton-like perturbations appear along the $x$- and $y$-axis. The solitary wave with local density minimum and a sharp phase gradient of the wave function at the position of the minimum, embedded in a 2D geometry leads to dominant decay mechanism. Thus, the soliton-like perturbations bend and decay into a more stable vortex-anti vortex pairs accompanied by sound waves. 
In our case $A$, we do not observe soliton-like perturbations! 
Furthermore, after applying these two methods ($A$ and $B$), the shape of the condensate is different, due to the fact that, in $A$ we do not have solitary waves.
\bigskip

Application of these two methods ($A$ and $B$) lead to different physical behaviour. For example, the number of total, positive and negative vortices as a function of time show different behaviour in $A$ and in $B$. With the method $A$ more negative vortices appear, which contribute to more annihilations and different total number of vortices as a function of time. Compare, for example these numbers of vortices, for the case when $\Omega$ was set to $0$ after the application of these methods. 
Our observation is that, the variation of different numbers of vortices with time vary more slowly in the case $A$.
\bigskip

We conclude that, in case $A$ the complex conjugation method reverses the rotation of the vortices in the two halves of the box. Now, the system contains roughly the same number of vortices with different sign and with different energies due to the corresponding different velocity fields. Thus, the imitation of the solid body rotation becomes damaged (stop of solid body rotation and far field almost cancels). 
\bigskip

The phase across the whole condensate follows the sign of the vortices, and becomes different in the upper and lower parts of the box. At the centre of the box, the phase is zero and by that, no phase kinks happen upon complex conjugation. Thus, no soliton like perturbations form. Very different energies appear and contribute to difficulty in experimental utilization. We do not know a way to realize this method experimentally.
\bigskip

Our study about a more global properties of the phase across the whole condensate shows that, this new method of complex conjugation change the rotation of the vortices in the upper half part of the box. On the other hand, with the phase imprinting method the system try to smooth out the change caused by generation of a discontinuity in the phase in the form of solitary and sound waves.

\section{Acknowledgements} 
The author is very grateful to Carlo F. Barenghi and Vicente Delgado for suggestions and discussions.



\begin{thebibliography}{99}
\bibitem{Kolmogorov}
A. N. Kolmogorov,
Dokl. Akad. Nauk. SSSR {\bf 30} 301 (1941);
Proc. R. Soc. London, Ser. A {\bf 434} 9 (1991).
\bibitem{Nore}
C. Nore, M. Abid, M. E. Brachet,
Phys. Rev.Lett. {\bf 78}, 3896 (1997).
\bibitem{Kobayashi}
M. Kobayashi and M. Tsubota,
Phys. Rev. Lett. {\bf 94} 065302 (2005).
\bibitem{Parker}
N. G. Parker and C. S. Adams,
Phys. Rev. Lett. {\bf 95} 145301 (2005).
\bibitem{Onsager}
L. Onsager,
Nuovo Cimento Suppl.{\bf 6} 249 (1949).
\bibitem{Feynman}
R. P. Feynman,
{\it Progress in Low Temperature Physics}, edited by C. J. Gorter
(North-Holland, Amsterdam)  (1955).
\bibitem{Vinen}
W. F. Vinen and J. J. Niemela,
J. Low Temp. Phys. {\bf 128} 167 (2002).
\bibitem{Donnelly}
R. J. Donnelly,
{\it Quantized Vortices in Helium II}.
Cambridge University Press, Cambridge,1991.
\bibitem{Saffman}
P. G. Saffman,
{\it Vortex Dynamics}.
Cambridge University Press, Cambridge,1993.
\bibitem{Yarmchu}
E. J. Yarmchu and R. E. Packard,
J. Low Temp. Phys. {\bf 46} 479 (1982).
\bibitem{Vinen1}
W. F. Vinen,
Proc. R. Soc. London, Ser A {\bf 240} 114 (1957);
{\bf240} 128 (1957); {\bf 240} 493 (1957).
\bibitem{Schwarz}
K. W. Schwarz,
Phys. Rev. B {\bf 31} 5782 (1985);
{\bf38} 2398 (1988).
\bibitem{Barenghi}
C. F. Barenghi, R. J. Donnelly, and W. F. Vinen,
{\it Quantized Vortex Dynamics and Superfluid Turbulence}.
Springer, 2001.
\bibitem{Davis}
S. I. Davis, P. C. Hendry, and P. V. E. McClintock,
Physica B {\bf 280} 43 (2000).
\bibitem{Fisher}
S. N. Fisher, A. J. Hale, A. M. Gu\'enault, and G. R. Pickett,
Phys. Rev. Lett. {\bf 86} 244 (2001).
\bibitem{Madison}
K. W. Madison, F. Chevy, W. Wohlleben, and J. Dalibard,
Phys. Rev. Lett. {\bf 84} 806 (2000).
\bibitem{Abo-Shaeer}
J. R. Abo-Shaeer, C. Raman, J. M. Vogels, and W. Ketterle,
Science {\bf292} 476 (2001).
\bibitem{Kasamatsu}
K. Kasamatsu, M. Tsubota, and M. Ueda,
Phys. Rev. A {\bf67} 033610 (2003).
\bibitem{Stalp}
S. R. Stalp, L. Skrbek, and R. J. Donnelly,
Phys. Rev. Lett. {\bf 82} 4831 (1999).
\bibitem{Finne}
A. P. Finne, T. Araki, R. Blaauwgeers, V. B. Eltsov, N. B. Kopnin, M. Krusius, L. Skrbek, M. Tsubota, and G. E. Volovik,
Nature (London) {\bf 424} 1022 (2003).
\bibitem{Kobayashi1}
M. Kobayashi and M. Tsubota,
Phys. Rev. A {\bf76} 045603 (2007).
\bibitem{Regnault}
N. Regnault and T. Jolicoeur,
Phys. Rev. B {\bf69} 235309 (2004).
\bibitem{Tsubota}
M. Tsubota, K. Kasamatsu, and M. Ueda,
Phys. Rev. A {\bf65} 023603 (2002).
\bibitem{Madarassy}
E. J. M. Madarassy and C. F. Barenghi,
J. Low Temp. Phys. {\bf 152} 122-135 (2008).
\bibitem{Munoz}
A. Munoz Mateo, and V. Delgado,
Phys. Rev. A {\bf74} 065602 (2006).
\bibitem{Munoz1}
A. Munoz Mateo, and V. Delgado,
Phys. Rev. A {\bf75} 063610 (2007).
\bibitem{Munoz2}
A. Munoz Mateo, and V. Delgado,
Phys. Rev. A {\bf77} 013617 (2008).
\bibitem{Ruprecht}
P. A. Ruprecht, M. J. Holland, K. Burnett, and M. Edwards,
Phys. Rev. A {\bf51} 4704 (1995).
\bibitem{Wang}
S. Wang, Y. A. Sergeev, C. F. Barenghi, and M. A. Harrison,
J. Low Temp. Phys. {\bf 149} 65 (2007).
\bibitem{Madarassy1}
E. J. M. Madarassy and C. F. Barenghi,
Geophysical and Astrophysical Fluid Dynamics {\bf103} 269-278 (2009).
\bibitem{Feder}
D. L. Feder, M. S. Pindzola, L. A. Collins, B. I. Schneider, and C. W. Clark,
Phys. Rev. A {\bf62} 053606 (2000).
\bibitem{Madarassy2}
E. J. M. Madarassy,
{\it Decay of soliton-like perturbations into vortex - anti vortex pairs}.
Accepted by the Romanian Journal of Physics (2009).

\end{thebibliography}
\end{document}